\documentclass[%
 reprint,
 amsmath,amssymb,
 aps,
]{revtex4-2}

\usepackage{graphicx}
\usepackage{dcolumn}
\usepackage{bm}

\usepackage{algpseudocode}
\usepackage{amssymb}
\usepackage{amsmath}
\usepackage{booktabs}
\usepackage{arydshln}
\usepackage{multirow}
\usepackage{hyperref}

\usepackage{xcolor}

\newcounter{algorithm}
\renewcommand{\thealgorithm}{\arabic{algorithm}}

\makeatletter
\newcommand{\algcaption}[1]{%
  \refstepcounter{algorithm}%
  \par\medskip
  \noindent\hrulefill\par
  \noindent\textbf{Algorithm \thealgorithm.} #1\par
  \noindent\hrulefill\par\smallskip
}
\newenvironment{breakablealgorithm}
  {%
   \begin{center}
   \begin{minipage}{0.95\columnwidth}
  }
  {%
   \par\smallskip\noindent\hrulefill
   \end{minipage}
   \end{center}
   \medskip
  }
\makeatother

\begin{document}
\title{From membership privacy leakage to Quantum Machine Unlearning} 

\author{Junjian Su$^{1,2}$, Runze He$^{1}$, Guanghui Li$^{1}$, Sujuan Qin$^{1}$, Zhimin He$^{3}$, Haozhen Situ$^{4}$, Fei Gao$^{1,2,}$} \email{gaof@bupt.edu.cn}

\affiliation{$^1$State Key Laboratory of Networking and Switching Technology, Beijing University of Posts and Telecommunications, Beijing 100876, China}
\affiliation{$^2$State Key Laboratory of Cryptology, P.O. Box 5159, Beijing 100878, China}
\affiliation{$^3$School of Electronic and Information Engineering, Foshan University, Foshan 528000, China}
\affiliation{$^4$College of Mathematics and Informatics, South China Agricultural University, Guangzhou 510642, China}


\date{\today}

\begin{abstract}
Quantum Machine Learning (QML) has the potential to achieve quantum advantage for specific tasks by combining quantum computation with classical Machine Learning (ML).
In classical ML, a significant challenge is membership privacy leakage, whereby an attacker can infer from model outputs whether specific data were used in training.
When specific data are required to be withdrawn, removing their influence from the trained model becomes necessary.
Machine Unlearning (MU) addresses this issue by enabling the model to forget the withdrawn data, thereby preventing membership privacy leakage.
However, this leakage remains underexplored in QML.
This raises two research questions: do QML models leak membership privacy about their training data, and can MU methods efficiently mitigate such leakage in QML models?
We investigate these questions using two Quantum Neural Network (QNN) architectures, a basic QNN and a Hybrid QNN, evaluated in noiseless simulations and cloud quantum device demonstrations.
To answer the first question, we analyze how quantum constraints shape membership privacy leakage in QML and then formalize a realistic gray-box threat model accordingly. Based on this, we design a membership inference attack (MIA) tailored to QNN outputs, and our results provide clear evidence of membership leakage in both QNNs.
To answer the second question, we propose a Quantum Machine Unlearning (QMU) framework, comprising three MU mechanisms. 
Evaluations on two QNN architectures show that QMU removes the influence of the withdrawn data while preserving accuracy on retained data. 
A comparative analysis further characterizes the three MU mechanisms with respect to data dependence, computational cost, and robustness.
We further study how the shot count in quantum measurement affects both membership leakage and unlearning stability.
Overall, this work provides a potential path towards privacy-preserving QML.
\end{abstract}

\maketitle


\section{INTRODUCTION}\label{sec:intro}

Leveraging unique physical phenomena such as quantum superposition, quantum computing exhibits exceptional potential in tackling high-dimensional and multivariate problems far beyond the reach of classical computing \cite{preskill18,arute19}.
Quantum Machine Learning (QML) integrates quantum computation with classical Machine Learning (ML), which allows for data representation within high-dimensional Hilbert spaces \cite{biamonte17,schuld15}. This capability positions QML as particularly well-suited for tasks characterized by stringent computational complexity requirements.
In recent years, QML has shown promising potential across a range of applications, including chemistry \cite{peruzzo14,cirstoiu20}, combinatorial optimization problems \cite{ni24, farhi14,zhao24,li25}, data analysis \cite{aimeur07, farhi18, rebentrost14}, quantum error correction \cite{sivak23,nautrup19}, and related areas.

The rapid advancement of QML technologies has sharpened attention to its security and privacy issues. 
On the one hand, QML inherits many known vulnerabilities from classical ML, such as adversarial attacks \cite{west23,liao21}, data-poisoning attacks \cite{kundu24}. 
On the other hand, QML confronts novel attack surfaces unique to quantum systems, such as interference-based attacks on quantum states \cite{zurek03}, and manipulation or spoofing of quantum algorithms outcomes \cite{franco24}. 
Meanwhile, several preliminary defense mechanisms have been proposed.
These include enhancing model robustness by exploiting quantum hardware noise or the unpredictability of superposition \cite{gong24}, and developing secure communication protocols to support cross-device model training \cite{alhashim2025}.

Although QML security has been explored from multiple perspectives, there remains no systematic approach to explore membership privacy leakage and its mitigation.
In classical ML, a key issue is the leakage of membership privacy, where an attacker can deduce whether specific data was involved in the model's training process based on its outputs \cite{shokri17, zhang21}.
This issue is further emphasized by major global data protection regulations, which explicitly mandate the principles of data withdrawal and the right to be forgotten \cite{calzada22, gdpr18, cppa22}.
When data owners request the withdrawal of specific data, it is essential to eliminate the influence of the removed data from the trained model.
However, retraining a model from scratch without the withdrawn data is impractical due to computational effort.
To address this, Machine Unlearning (MU) has emerged as an essential privacy-preserving technique, which enables a trained model to behave as if withdrawn data had not been used \cite{bourtoule21, trippa24}.
Despite the critical role of membership privacy leakage and MU algorithms, studying these issues in QML remains an open problem.

This paper explores two core questions: do QML models leak membership privacy about training data, and can MU methods efficiently mitigate this leakage risk in QML models?
To address these questions, we evaluate two types of QNN, specifically a basic Quantum Neural Network (basic QNN) and a Hybrid QNN (HQNN), both using hardware-efficient ansatz with a 5-layer depth, in both noiseless simulations and cloud quantum device demonstrations.
To address the first question, we analyze QML-specific sources of membership privacy leakage. Because intermediate quantum states cannot be inspected without measurement-induced disturbance (which alters subsequent evolution), and unknown states cannot be perfectly cloned, classical white-box assumptions are impractical for deployed QNN services. Accordingly, we formalize a realistic gray-box inference-API threat model in which the adversary can only query the deployed QNN.
Based on this threat model, we design a Membership Inference Attack (MIA) tailored to QNN to infer membership status.
Our evaluation results clearly demonstrate that both types of QNN models exhibit measurable membership privacy leakage in simulation and on a cloud quantum device.
For the second question, we propose Quantum Machine Unlearning (QMU), which integrates three distinct MU mechanisms.
Our results validate the effectiveness of QMU on two types of QNN, demonstrating that it successfully removes the influence of withdrawn data while preserving accuracy on the retained dataset.
A subsequent comparative analysis further reveals that the three MU mechanisms exhibit distinct tradeoffs in data dependence, computational cost, and robustness.
Numerical simulations show that an appropriately low shot count can mask membership signatures with only a minor impact on accuracy, benefiting the provider.  
By contrast, a higher shot count improves both attack reliability and unlearning efficacy, benefiting the attacker and withdrawing users. 
Accordingly, we advocate a phase-dependent shot configuration: high shots for maintenance (training/unlearning) and low shots for deployment (inference APIs).
Overall, this work investigates membership privacy across two QNN types by demonstrating leakage risk and proposing QMU to mitigate it, thereby paving a potential path toward developing more secure QML.

The remainder of this paper is organized as follows. Section \ref{sec:RE} reviews related work in QML and in MU for classical ML. Section \ref{SEC_MIA} addresses our first research question by detailing the MIA methodology and presenting evidence of membership privacy leakage in QML models. In Section \ref{sec3}, we introduce the QMU framework, present three MU mechanisms, and evaluate their effectiveness. Finally, Section \ref{sec-con} concludes with a summary of our findings and future research directions.

\section{RELATED WORK}\label{sec:RE}
This section reviews two research areas that are highly relevant to this study: 
(1) the development and modeling paradigms of Quantum Machine Learning (QML), and (2) the emerging field of Machine Unlearning (MU) in classical Machine Learning (ML). 
By examining the progress made in these two directions, we aim to highlight the technical challenges and research gaps surrounding adversarial attacks and defense mechanisms related to membership privacy in QML models.

\subsection{Quantum machine learning}
Classical ML has achieved significant breakthroughs in complex tasks. However, it now faces a new set of computational challenges \cite{kaplan20}. 
First, the increasing complexity of model architectures necessitates a growing reliance on substantial computational resources during training \cite{thompson20}. 
Second, data are becoming increasingly high-dimensional and dynamically evolving, rendering traditional methods progressively inadequate for representing complex features and modeling nonlinear patterns \cite{fan14}.
Against this backdrop, quantum computing offers a new computational paradigm for ML by leveraging intrinsic parallelism and high-dimensional Hilbert-space representations \cite{buhlmann11}. Consequently, QML integrates quantum computation with classical ML to pursue solutions to computationally demanding learning problems \cite{preskill18,biamonte17,chen25,liling25}.

Early QML work largely lifted classical algorithms into the quantum domain. Representative examples include Rebentrost’s quantum support vector machine (QSVM), built on the HHL routine \cite{rebentrost14}, and quantum principal component analysis (QPCA) and quantum clustering \cite{lloyd14}.

Although these methods theoretically established the potential for quantum speedup, their reliance on idealized assumptions about quantum states often rendered them challenging to implement on contemporary quantum hardware. 
Subsequently, the introduction of Parameterized Quantum Circuits (PQCs), which provide enhanced flexibility for modeling complex quantum states, drove the widespread adoption of quantum–classical hybrid architectures \cite{benedetti19}. 
The capacity of PQCs to improve generalization and stability, particularly for datasets that are both high-dimensional and small-sample, has spurred significant research interest in their optimal structure and optimization \cite{su25, he24, li25zx}. 
This developmental period saw the emergence of systematic QML modeling frameworks, including quantum kernel methods \cite{schuld21}, Variational Quantum Classifiers \cite{li22}, and Quantum Circuit Learning (QCL) \cite{mitarai18}. 
QML has thus entered the noisy intermediate-scale quantum (NISQ) era, marking a crucial empirical phase in the pursuit of quantum advantage \cite{arute19}.

\subsection{Machine unlearning}
Driven by growing imperatives to uphold data sovereignty and user privacy, data revocability has emerged as a critical requirement for the compliant design of ML models. 
Storage-level deletion is insufficient because trained models can retain statistical traces of removed samples. 
However, trained models often retain statistical traces of sensitive samples, failing to achieve genuine data erasure.
Given the prohibitive computational expense associated with retraining models from scratch, researchers have increasingly prioritized the development of more efficient Machine Unlearning (MU) approaches. 
These methods are specifically engineered to selectively eliminate the influence of specified data points on a trained model while rigorously preserving its overall performance.

MU tasks are typically categorized into three fundamental types: class unlearning, instance unlearning, and feature unlearning \cite{bourtoule21, trippa24}.
Class unlearning involves removing all samples belonging to a particular category and subsequently adapting the model so that it no longer possesses recognition capability for that class. 
Instance unlearning aims to precisely eliminate the statistical influence of a specific data sample, requiring the model to behave as though the sample had never been included in the training set.
Feature unlearning focuses on reducing the model’s dependency on specific sensitive attributes, such as gender or age, primarily to mitigate inherent bias or sensitivity issues.

To address various types of unlearning tasks, mainstream MU approaches are generally classified into two overarching categories: data reorganization and model manipulation methods. 
Data reorganization methods strategically partition the training data into structured subsets to enable localized retraining, thereby facilitating efficient unlearning. 
Model manipulation methods directly alter the model parameters using techniques such as influence functions, gradient ascent optimization, or pruning to effectively remove the impact of target samples.
To comprehensively evaluate the effectiveness of unlearning mechanisms, researchers have introduced various metrics, including MIA success rate, output divergence, and parameter perturbation \cite{bourtoule21}. 
The retained model performance is typically quantified using basic metrics like accuracy or F1 score.
In practical applications, additional considerations include computational overhead, data dependency, and scalability of the unlearning method.

While finalizing this manuscript, we noted the contemporaneous work by Zhang et al. \cite{chenzhang25}. 
Their study introduces MU to QNN to address data-poisoning attacks within a binary classification setting, where they compare the ability of Multi-Layer Perceptrons (MLP) and QNN to unlearn corrupted samples. In contrast, our work focuses rigorously on the pervasive problem of membership privacy leakage. We propose and validate the comprehensive QMU framework, which we evaluate on ten classification tasks across both noiseless simulations and a cloud quantum device, demonstrating a broader scope in terms of problem domain and empirical validation environment.

\section{MEMBERSHIP PRIVACY LEAKAGE IN QML}
\label{SEC_MIA}
\subsection{Methods}

\begin{figure*} 
  \centering
  \includegraphics[width=0.95\textwidth]{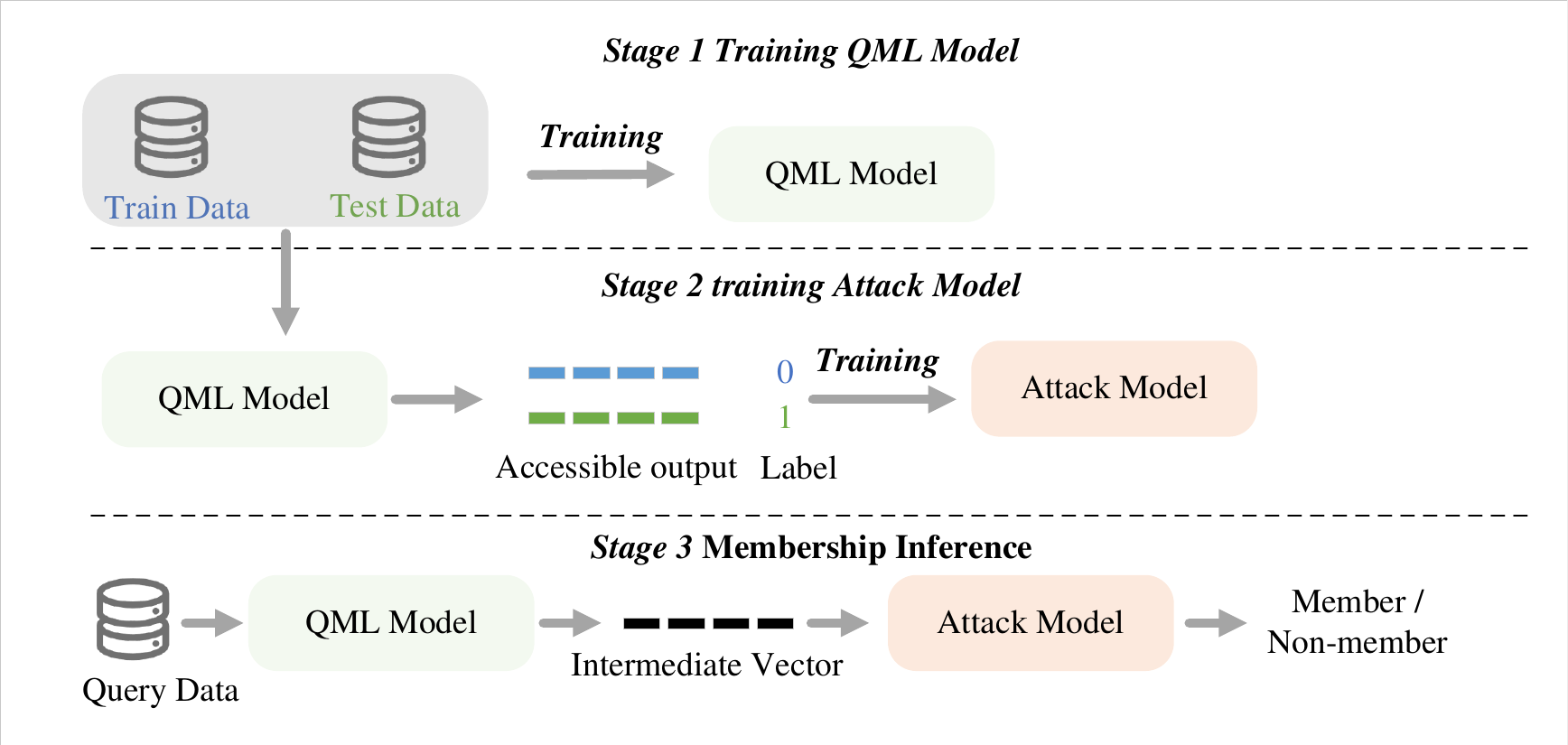} 
  \caption{Membership Inference Attack Workflow on QML Models.
  (Stage 1) the initial training of the target QML model; (Stage 2) training the attack model using accessible output obtained via an inference API, by querying the target model on known member and non-member data; and (Stage 3) the final inference step to predict a query sample’s membership status.}
  \label{fig:MIA}
\end{figure*}

This section systematically investigates membership privacy leakage in trained QML models to answer the core question: Do QML models leak membership information about their training data?
Motivated by QML-specific constraints relevant to privacy leakage, we adopt a realistic gray-box threat model and develop a MIA tailored to QNN.
The workflow is illustrated in FIG. \ref{fig:MIA}. This section is organized as follows: we describe the target QML methodology (Stage 1), followed by the threat model and the MIA procedure (Stages 2--3).
For clarity, we distinguish throughout this work between noiseless simulations and cloud quantum device demonstrations performed through cloud access to a superconducting quantum processor. The cloud quantum device part of this study is not a traditional laboratory experiment on a custom-built device; rather, it is a reproducible cloud-quantum-computing demonstration carried out under the provider's execution stack, shot budget, and recorded calibration state.

\subsubsection{Quantum machine learning}
\label{sec3_QML}
To systematically evaluate the behavior of QML models when subjected to privacy attacks, this study adopts the Quantum Neural Network (QNN) as the representative model. 
QNNs are particularly well-suited to the constraints of current Noisy Intermediate-Scale Quantum (NISQ) devices due to their strong expressive power and training flexibility, which have led to their widespread use in small- to medium-scale quantum classification tasks. 
The following section provides a detailed technical description of the QNN architecture and the training procedure utilized in this study.

\begin{figure*}
    \centering
    \includegraphics[width=0.8\textwidth]{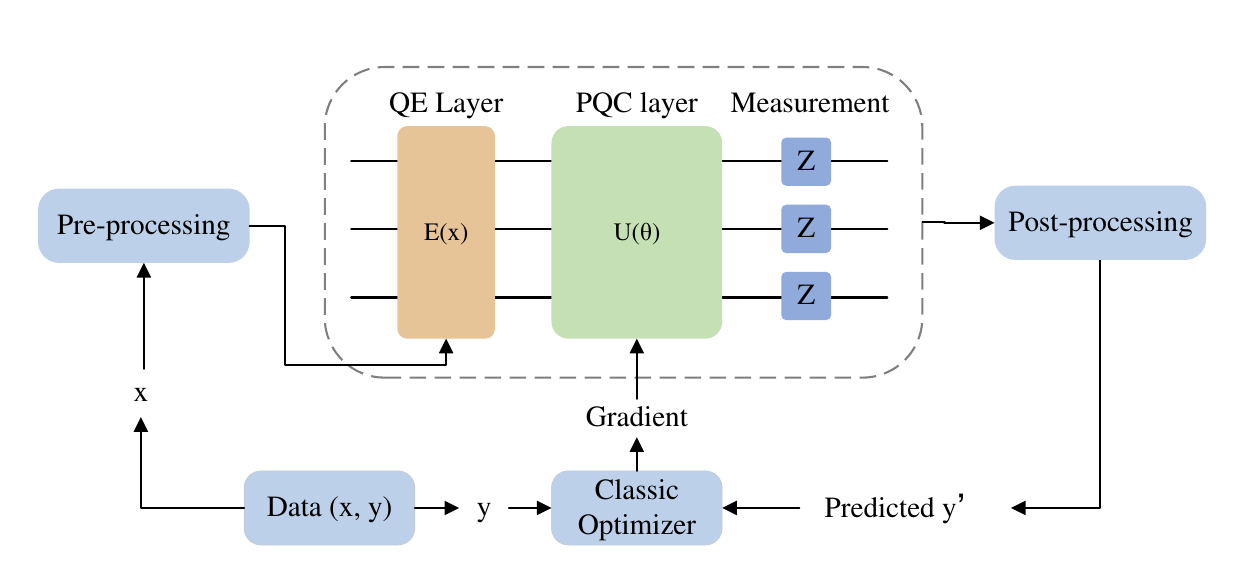}
    \caption{Architecture and data flow of the QNN model. 
The model consists of a quantum encoding layer $E(x)$ that encodes classical input $x$ into a quantum state, 
a PQC layer $U(\theta)$ that evolves this quantum state based on trainable parameters, and a measurement layer that performs Pauli-Z measurements to extract classical observables.
Classical pre-processing is employed to reduce the input dimension to match the available number of qubits, while post-processing maps the measurement results to the final prediction space. 
During the training phase, the predicted output $y'$ is compared with the ground truth label $y$ to compute the loss, which subsequently guides the iterative parameter updates via gradient descent.}
\label{fig:QNN}
\end{figure*}

As illustrated in FIG.\ref{fig:QNN}, the QNN architecture is fundamentally composed of three core components: a Quantum Encoding (QE) layer, a PQC layer, and a Measurement layer. The Quantum Encoding Layer is responsible for transforming the classical input data $x$ into a quantum state:
\begin{equation}
|\psi_{(x)}\rangle=E(x)|0\rangle,
\label{eqn:encoder}
\end{equation}
where $E(x)$ represents a specific encoding transformation, which may include techniques such as angle encoding, amplitude encoding, or phase encoding.
The PQC layer, also referred to as the Ansatz, is subsequently utilized to parameterize the evolution of the quantum state $|\psi(x)\rangle$ within the high-dimensional Hilbert space, thereby capturing intrinsic data features. 
This evolutionary process can be formally expressed as:
\begin{equation}
|\psi_{(x,\theta)}\rangle = U(\theta)|\psi_{(x)}\rangle,
\label{eqn:quantum_circuit}
\end{equation}
where $U(\theta)=\prod_{\ell=1}^{L} U_{\ell}(\theta_{\ell})$ denotes an $L$-layer PQC, and $\theta=\{\theta_{\ell}\}_{\ell=1}^{L}$ the trainable parameters.
The specific structure of $U$ includes a general hardware-efficient structure as well as a specialized Ansatz designed based on the adaptive and quantum architecture search algorithms \cite{tang21, situ24, he22, zhang22, du22, he25}. 
Finally, the Measurement Layer is responsible for performing measurement operations on the evolved quantum state, thereby extracting the classical observables:
\begin{equation}
m_i(x)=\langle \psi_{(x,\theta)} \mid Z_i \mid \psi_{(x,\theta)} \rangle,\quad i=1,\ldots,n,
\label{eqn:Mlayer}
\end{equation}
where $m_i(x)\in[-1,1]$ denotes the Pauli-$Z$ expectation value on qubit $i$, and $\mathbf{m}(x)=(m_1(x),\ldots,m_n(x))$ provides the measurement features used for downstream classification.
In addition to the above fundamental components, QNNs may incorporate a Pre-processing Layer and a Post-processing Layer to accommodate hardware limitations and practical task requirements.
Owing to the limited number of available qubits in current quantum processors, the direct encoding of high-dimensional classical data is often infeasible. 
Therefore, dimensionality reduction techniques, such as Principal Component Analysis (PCA) or neural network-based feature compression, are frequently applied during pre-processing to reduce the input dimensionality to a level suitable for quantum encoding.
Moreover, the Post-processing Layer maps the measurement results to the dimension required for the target prediction task. A Multi-Layer Perceptron (MLP) is commonly used as the post-processing layer, consisting of a hidden layer and an activation layer. 
The hidden layer maps the measurement feature vector $\mathbf{m}(x)$ to logits
\begin{equation}
\mathbf{z}(x)=W\,\mathbf{m}(x)+\mathbf{b}, \qquad W\in\mathbb{R}^{K\times n},\ \mathbf{b}\in\mathbb{R}^{K},
\end{equation}
where $W$ and $\mathbf{b}$ are trainable parameters in the MLP, and $K$ is the number of classes. The logit vector $z(x)$ represents the raw, unnormalized scores for each class. The activation layer then applies the softmax function to produce the class-probability vector $\boldsymbol{\pi}(x)$ via
\begin{equation}
\pi_k(x) = \frac{\exp(z_k(x))}{\sum_{j=1}^{K} \exp(z_j(x))}, \quad k = 1, \ldots, K.
\end{equation}
The class-probability vector $\boldsymbol{\pi}(x)$ represents the predicted probability distribution over the $K$ classes, with each component $\pi_k(x)$ corresponding to the probability that the input $x$ belongs to class $k$.

The training process of QNN generally follows these steps. 
(1) Constructing the dataset $(x, y)$ and initializing the parameters $\theta$; 
(2) Performing a forward pass to obtain class probabilities $\boldsymbol{\pi}(x)$.
(3) Calculating loss using the cross-entropy loss function (in the case of classification tasks):
\begin{equation}
\mathcal{L} = -\sum_{i=1}^{K} y_i \log(\boldsymbol{\pi}_i(x)),
\label{eqn:Loss}
\end{equation}
where $K$ represents the number of classes. 
(4) The gradient of the parameters is computed based on the loss function, and the parameters $\theta$ are updated:
\begin{equation}
          \theta^{k+1}
          = \theta^{k}
          - \eta \,\nabla_{\theta^{k}}\mathcal{L}\bigl(f_{\theta^{k}}(x_t),\, y_t\bigr),
\label{GD}
\end{equation}
where $\theta^{k+1}$ represents the updated parameters at iteration $k+1$, and $x_t$ and $y_t$ are the $t$-th sample from the dataset $D$, with $\eta$ denoting the learning rate.
(5) Repeat Steps (2) through (4) until the model satisfies a predefined convergence criterion.
The QNN model, defined by this architecture and training process, is designed to effectively harness the advantages of quantum computation while fully accommodating the practical constraints of current NISQ devices. 
This architecture provides a representative foundation for the subsequent privacy-attack and unlearning evaluation.

\subsubsection{Threat Model and membership inference attacks}
\label{sec3_PA}

In QML, privacy leakage analysis must account for the informational constraints imposed by quantum mechanics. 
In classical ML, white-box attacks typically assume access to internal model information, including intermediate representations, without disturbing the subsequent computation.
However, this assumption does not hold in QML.
First, due to measurement-induced collapse, any attempt to access intermediate quantum information requires a measurement, which projects the state and typically perturbs the subsequent evolution, potentially changing the final outcome.
Moreover, even if an adversary attempts to avoid disturbing a single run by duplicating the intermediate state for offline analysis, the no-cloning theorem forbids perfect copying of unknown quantum states.
As a result, the white-box capabilities commonly assumed in classical privacy attacks are not available by default in the QML service setting, unless one assumes unrealistic privileges to intervene in the quantum execution.

In this work, we adopt a gray-box inference-API threat model, in which the adversary can only query the target QNN and observe classical post-measurement outputs, without access to or control over the circuit implementation (e.g., topology and variational parameters) or the measurement procedure. 
Concretely, a single query to the QNN inference service returns raw bitstring counts under a fixed $N_{shots}$, from which observable expectation values $\mathbf{m}(x)$ are estimated as empirical averages.
The $\mathbf{m}(x)$ are then passed through classical post-processing layers to produce logits and corresponding softmax probabilities. If the true label y is known to the adversary, the per-sample cross-entropy loss can be computed locally from the predicted probabilities.
Accordingly, we restrict the attacker’s accessible output features for training the attack model to the following:
(1) observable expectation values $\mathbf{m}(x)$,
(2) logits obtained after classical post-processing $\mathbf{z}(x)$,
(3) softmax probability vectors $\pi_k(x)$, and
(4) loss values $\mathcal{L}$ computed from predicted probabilities and known labels.
Importantly, as quantum measurements obey the Born rule, the outputs of a QNN are subject to inherent shot noise due to finite sampling uncertainty. 
The measured value of a quantum observable for a specific input \(x\) is given by:
\begin{equation}
\hat{m}_i(x) = \frac{1}{N_{shots}} \sum_{s=1}^{N_{shots}} (-1)^{b_i^{(s)}},
\label{eqn:shotNoise}
\end{equation}
where \(b_i^{(s)} \in \{0,1\}\) denotes the individual measurement outcome for the \(i\)-th feature and \(N_{shots}\) represents the number of measurements performed. 
This shot noise introduces fluctuations around the true expectation value (as shown in Eq.\ref{eqn:Mlayer}), and becomes larger as \(N_{shots}\) decreases.
Prior work has suggested that this shot noise can be leveraged to achieve quantum differential privacy (QDP) \cite{hirche23qdp, zhouying17qdp,li24qdp_projection}.
To investigate the impact of shot noise on MIA, we impose a strict per-sample query budget, limiting the adversary to a single query per sample and thereby preventing repeated querying from averaging out shot noise. We do not claim a formal QDP guarantee; rather, we only show that finite-shot noise acts as a privacy-relevant physical factor under this threat model.
Our next evaluations are designed to systematically evaluate how varying shot counts and different feature types will affect MIA performance.

Under the gray-box inference-API threat model, we employ MIAs to quantify membership privacy leakage in QML (Stages 2--3 in FIG. \ref{fig:MIA}). The overall procedure is summarized in Algorithm \ref{alg:MIA}. 
First, to construct the training set for the attack model, the adversary queries the target QNN on known members \(D_{\text{in}}\) and non-members \(D_{\text{out}}\), and extracts the output features \(\phi(x)\), which may consist of the measurement outcomes $\mathbf{m}(x)$, the post-processing logits $\mathbf{z}(x)$, the predicted probabilities $\pi(x)$ or $\pi_k(x)$, and the loss value $\mathcal{L}(x,y)$.
Second, the attack model is trained on the resulting labeled pairs \((\text{features},\, \text{member/non-member})\). 
Third, the trained attack model is applied to candidate samples to predict their membership status, and its performance is used to characterize the leakage intensity.
Finally, given a query sample $x_{\text{query}}$, the trained attack model takes the corresponding output features $\phi(x_{\text{query}})$ from the target model, outputs the score $\text{MLP}(\phi(x_{\text{query}}))$, and predicts the sample as a member if the score is greater than $0.5$, and as a non-member otherwise.
As specified in Algorithm \ref{alg:MIA}, we adopt a worst-case evaluation protocol: the adversary is assumed to have access to labeled member and non-member examples with respect to the target QNN. 
This assumption is used to estimate an upper bound on privacy leakage and does not imply that a black-box API user in deployment can directly obtain the model owner's private training set. 
In realistic scenarios, such labeled examples may arise in withdrawing users or data-contributor settings, or can be approximated via standard shadow-model strategies using auxiliary data drawn from a similar distribution.

\begin{breakablealgorithm}
\algcaption{Membership Inference Attack Process}\label{alg:MIA}
\begin{algorithmic}[1]
\State \textbf{Input:} target QNN model $A$, known member dataset $D_{\mathrm{in}}$, non-member dataset $D_{\mathrm{out}}$
\State \textbf{Output:} MIA success rate
\For{each sample $x \in D_{\mathrm{in}} \cup D_{\mathrm{out}}$}
    \State Query the QNN $A$ to obtain output features $\phi(x)$
    \State Store labeled pair $(\phi(x), \mathrm{member/nonmember})$ in training set
    \State $\mathcal{D}_{\mathrm{attack}}$
\EndFor
\State Train an attack model on $\mathcal{D}_{\mathrm{attack}}$
\For{each query sample $x_{\mathrm{query}}$}
    \State Obtain output features $\phi(x_{\mathrm{query}})$ from $A$
    \State Predict the query sample as a member if
    \State $MLP(\phi(x_{\mathrm{query}})) > 0.5$, and as a non-member otherwise
\EndFor
\State Compute MIA success rate based on correct predictions of membership
\end{algorithmic}
\end{breakablealgorithm}

\subsection{Simulation and cloud quantum device demonstrations}

In this section, we investigate whether unprotected QNN models exhibit membership-privacy leakage under realistic adversarial access. 
We quantify leakage using the success rate of an MIA, which measures an adversary’s ability to decide whether a query sample was included in the target model’s training set based solely on the model’s observable outputs. 
Our evaluation proceeds in three parts. 
First, we describe the evaluation setup, including the dataset and class-wise unlearning protocol, the two target architectures, and the execution environments, together with training and attack-model configurations. 
Second, we report MIA performance using different accessible output features (observable expectation values, logits, softmax probability vectors, loss values) to compare the leakage of the original model $A_o$ against the retrained baseline $A_t$. 
Third, we study the effect of finite-shot measurement by sweeping the inference shot count under a strict per-sample query budget, and we track how both MIA success rate (privacy) and classification accuracy (utility) vary with the shot budget. 
Overall, this protocol reflects practical deployment scenarios in which the model is accessed locally or via an inference API, and only classical, post-measurement outputs are observable by the adversary.

\subsubsection{Setup}

All results in this paper were conducted on the 10-class MNIST digit classification task under a class-wise unlearning paradigm, utilizing both noiseless simulations and a cloud quantum device. 
For the unlearning task, one class of data was randomly selected from the digits \{4,5,8\} to form the unlearn data $D_u$, with the remaining data constituting the retained dataset $D_r$.
We conduct evaluations using two representative QNN models, a basic QNN and an HQNN.
The basic QNN architecture incorporates PCA-based pre-processing, and the PQC layer consists of a 10-qubit, 5-layer hardware-efficient ansatz. The post-processing for basic QNN is performed by a fully connected layer mapping to a 10-dimensional output. 
The HQNN utilizes a classical CNN for pre-processing, which includes 3×3 convolutions, pooling, and dense layers, and is coupled with a 10-qubit, 5-layer hardware-efficient ansatz. The post-processing for HQNN is the same as basic QNN.
The dataset consists of 1,000 randomly selected MNIST images, partitioned into a training set $D^{\text{train}}$ (600) and a test set $D^{\text{test}}$ (400).
During training, the basic QNN is optimized with a learning rate of 0.05 and a batch size of 32, while the HQNN is trained with a learning rate of 0.10 and a batch size of 8.
The MIA is then constructed using 100 samples from the unlearned class to test whether the prediction behaviors of $A_o$ and $A_t$ diverge significantly on these inputs.
The attack model itself is implemented as a deep neural network, comprising three fully-connected layers, ReLU activation functions, and dropout regularization (p=0.3). 
During training, the attack model is optimized with a learning rate of 0.01 and a batch size of 15.

To ensure statistical reliability, we report results averaged over 20 random seeds. 
Noiseless simulations were performed using the PennyLane and Qiskit simulators.
For the simulation in Sec. III.B, we use analytic evaluations (Infinite shots) for both QNN and the attacking model. 
Cloud quantum device demonstrations were conducted on the Tianyan-504 superconducting quantum processor with $N_{shots}$ =4096 per circuit evaluation. 
The backend returns raw measurement counts, which we directly report and use in our analysis. No error-mitigation technique was applied. For reproducibility, we record a calibration snapshot at execution time (Aug, 2025): average $T_1 = 57.86\,\mu\mathrm{s}$, average $T_2^{*} = 6.27\,\mu\mathrm{s}$, readout error 9.61\%, single-qubit gate error 0.88\%, and two-qubit gate error 8.55\% (Tianyan-504 device dashboard). We use the provider’s default compilation/execution pipeline (no customized transpilation or routing).

\subsubsection{Results}
We evaluate the membership privacy leakage of QNN models by conducting the MIA on MNIST classification tasks.
For comparative analysis, the original model $A_o$, which acts as the unprotected benchmark for privacy leakage, is trained on the complete $D^{\text{train}}$.
In contrast, the target model $A_t$, representing the ideal unlearned state, is obtained by retraining exclusively on the retained dataset $D_r$. 
The attack relies on extracting four types of model outputs as features: 
(1) softmax probabilities $p(y|\mathbf{x}) \in \mathbb{R}^{10}$, (2) logits $\mathbf{z}(\mathbf{x}) \in \mathbb{R}^{10}$, and (3) cross-entropy loss $L(\mathbf{x},y) \in \mathbb{R}$, (4) observable expectation values.
Finally, we assess the privacy leakage risk for both $A_o$ and $A_t$ across the basic QNN and HQNN architectures via the MIA success rate.

\begin{table}[htbp]
    \centering
    \caption{
        MIA success rates (\%) on QNN models using different model outputs. 
        For both QNN and HQNN architectures, MIA is conducted using loss values, logits, and softmax probabilities as input features to the attack model.‘exp-values’ denotes the observable expectation values estimated from 4096 shots.
    }
    \label{tab:mia_qnn_hqnn}
    \begin{tabular}{llcc}
        \toprule
        QNN & Outputs & Original Model & Target Model \\
        \midrule
        \multicolumn{4}{c}{Noiseless Simulation} \\
        \midrule
        \multirow{3}{*}{Basic QNN}
            & Loss    & 84.3  & 4.9 \\
            & Logit   & 83.6  & 11.7 \\
            & Softmax & 75.2  & 20.4 \\
        \midrule
        \multirow{3}{*}{HQNN}
            & Loss    & 98.0  & 0.0 \\
            & Logit   & 100.0 & 4.8 \\
            & Softmax & 100.0 & 7.0 \\
        \midrule
        \multicolumn{4}{c}{Cloud Quantum Device} \\
        \midrule
        Basic QNN & exp-values  & 67.1 & 12.8 \\
        HQNN & exp-values  & 83.5 & 6.4 \\

        \bottomrule
    \end{tabular}
\end{table}

Table \ref{tab:mia_qnn_hqnn} presents the MIA success rates across different model configurations and execution environments.
First, we analyze the influence of different output features on MIA success rates using a quantum simulator. 
For the basic QNN model, attackers achieved high success rates (84.3\% using loss values, 83.6\% using logits, and 75.2\% using softmax outputs), indicating clearly membership privacy leakage. 
In the ideal target QNN model, these rates substantially decreased to 4.9\%, 11.7\%, and 20.4\%, respectively.
The HQNN architecture demonstrated even greater vulnerability in its original form, with MIA success rate reaching 98.0\% (loss), 100.0\% (logits), and 100.0\% (softmax), revealing heightened privacy risks in this higher-capacity QNN model.
For the target HQNN model, success rates dropped markedly to 0.0\%, 4.8\%, and 7.0\%, confirming that more expressive models exhibit stronger data retention, leading to a greater initial privacy vulnerability.
These results demonstrate that the MIA method can reliably identify training data presence in unprotected QNN models, exposing leakage risk of membership privacy.

We also investigated the impact of using measurement results, which can be acquired from a quantum cloud platform, on the MIA success rate, utilizing cloud quantum device.
Although the presence of noise affects the observable expectation values and lowers the absolute attack accuracy, an adversary can still infer membership by exploiting the performance gap between the original and target models.
Furthermore, the environmental noise introduced in the cloud quantum device appears to narrow the differential gap in MIA success rates between the two models, suggesting a slightly increased difficulty for an adversary to distinguish between them based purely on noisy observable expectation values. 
Overall, the MIA achieved high average success rates of 90\% in noiseless simulations and 75.5\% on cloud quantum devices, empirically validating the existence of a verifiable membership privacy leakage risk.

\subsection{Analysis of the Impact of Shot Noise on MIA}
\label{sec:shot_noise_mia}

Unlike traditional deep learning models that typically yield deterministic probability vectors, QNNs are fundamentally distinct: their outputs depend on finite measurements of quantum states. This measurement process introduces inevitable shot noise, characterized by statistical errors scaling as $\mathcal{O}(1/\sqrt{N_{shots}})$. Given that MIA relies on exploiting subtle statistical traces in model outputs to infer membership, this intrinsic physical noise directly obfuscates the fine-grained features relied upon by attackers, thereby altering the landscape of privacy leakage. Consequently, investigating QML security and privacy necessitates considering shot noise not merely as an error source, but as a critical physical variable.

To investigate whether this physical noise constitutes a natural defense mechanism, we designed an analysis with varying measurement shots. 
We trained the HQNN in analytic mode (effectively infinite shots) and evaluated privacy leakage in a realistic inference-API setting where the adversary can only observe finite-shot outputs returned by the API. 
By varying shot count $N_{\text{shots}}\in\{16,64,256,1024,8192\}$, we quantify the privacy and utility trade-off across shot budgets, using the MIA success rate to measure membership leakage and the classification accuracy to measure model utility.
The threat model is defined as a strict gray-box scenario: the model provider (Defender) controls and sets $N_{shots}$, while the attacker has no access to the underlying noiseless probability distribution. Instead, the attacker can only obtain noisy measurement results under the current $N_{shots}$ setting to train their attack model. This setup faithfully simulates a realistic constrained scenario where an attacker accesses a model via a quantum cloud API.

\begin{figure}[t]
    \centering
    \includegraphics[width=\linewidth]{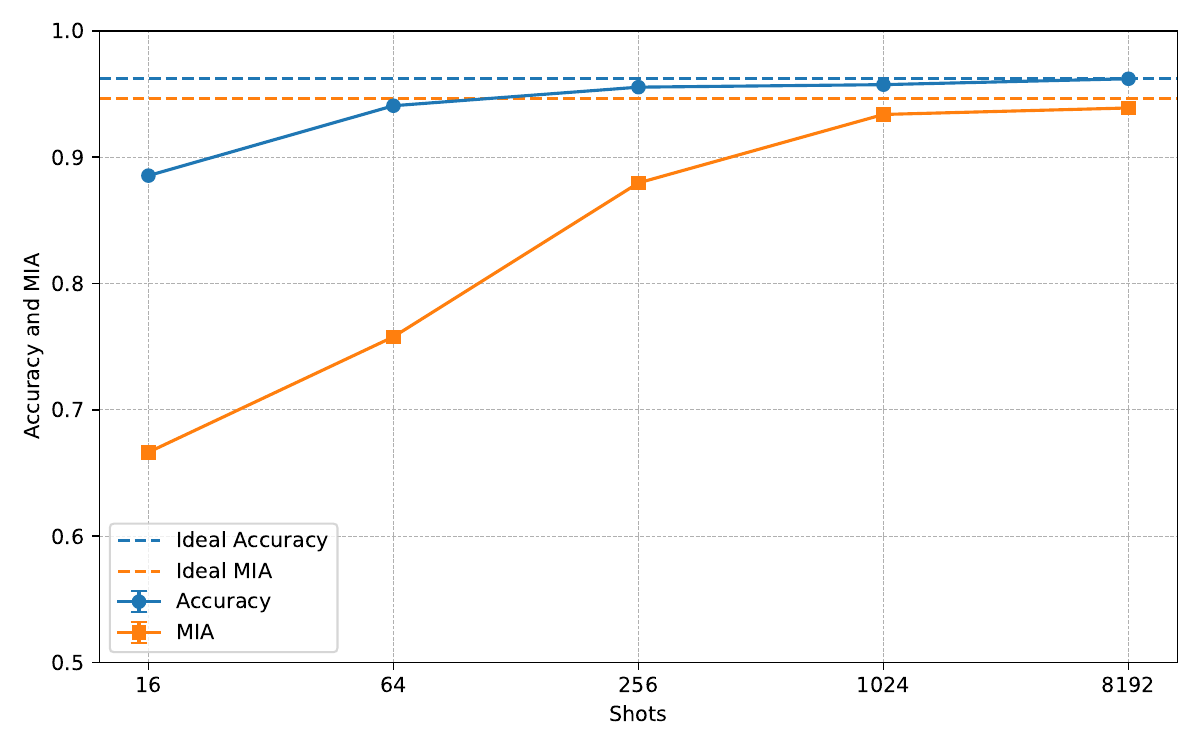} 
    \caption{
    Analysis of the impact of shot noise on model utility and privacy leakage. The horizontal axis represents shot count ($N_{shots}$), and the vertical axis denotes the percentage for both classification accuracy and MIA success rate. The blue and orange dashed lines represent the ideal noiseless accuracy and MIA success rate, respectively. As the number of shots decreases, the classification accuracy remains robust in the low-shot regime, while the MIA success rate decreases significantly. This suggests that intrinsic shot noise functions as a natural defense by masking the specific traces of training data.}
    \label{fig:shot_noise_mia}
\end{figure}

The numerical simulation results are presented in FIG. \ref{fig:shot_noise_mia}, which illustrates the dependence of both classification accuracy and MIA success rate on the number of measurement shots.
In the high-shot regime (e.g., $N_{shots}=8192$), where the measurement distribution approximates the exact expectation value, both the classification accuracy and MIA success rate of the HQNN converge to the ideal baselines. 
However, as the measurement count decreases to $N_{shots}=16$, the MIA success rate displays extreme sensitivity to noise, decreasing from $\sim 94\%$ to $\sim 67\%$; in contrast, the model's classification accuracy demonstrates remarkable robustness, decreasing only slightly from $\sim 96\%$ to $\sim 89\%$. 
This phenomenon, in which the MIA success rate drops sharply while accuracy declines only marginally, highlights the physical defense effect inherent in shot noise.
In low-shot regimes, significant statistical fluctuations preferentially mask the fragile discriminative information (or membership fingerprints) distinguishing members from non-members, thereby preventing attackers from extracting effective features from the model output. 
In contrast, the model’s classification accuracy is largely supported by more robust, decision-relevant statistics of the output distribution, which empirically appear less sensitive to shot noise than the fine-grained cues exploited by MIA.

In summary, model providers can actively tune shot count to exploit this physical noise as a foundational privacy defense. For instance, selecting 64 shots in our analysis effectively preserves model utility while simultaneously reducing the MIA success rate.
We emphasize that this effect is evaluated under a fixed per-sample query budget (or cost constraint): if an adversary can repeatedly query the same sample and average the outputs, the effective statistical noise can be reduced.
Moreover, even under this cost-limited setting, this passive defense remains imperfect: in the extreme case of $N_{shots}=16$, the MIA success rate ($68\%$) remains significantly above the random-guess baseline.

\subsection{Summary}
This section investigated the fundamental question of whether trained QNN models leak membership privacy concerning their training data.
By exploiting a comprehensive set of output features, MIA is able to reliably infer data's membership in both noiseless simulations and on a cloud quantum device. 
This phenomenon empirically confirms the existence of privacy leakage issues in QNN models. 
More broadly, our findings situate this work within the growing literature on quantum adversarial machine learning and extend it beyond robustness-centric adversarial settings to membership-privacy risks \cite{ren22eqal, lu20qaml}.
In response to our first core research question, we conclude that under current conditions, QNN models indeed present verifiable privacy leakage risks, and these risks can be systematically captured and quantified using feasible and practical attack strategies.
We further show that reducing $N_{shots}$ can partially suppress MIA performance under a single-query budget, but the leakage remains non-negligible. Therefore, to achieve thorough, precise, and lossless forgetting of specific sensitive data, we introduce the active QMU framework in Section \ref{sec3}.

\section{QUANTUM MACHINE UNLEARNING}
\label{sec3}

\subsection{Methods}

The previous section demonstrated that trained QNN models are vulnerable to membership privacy leakage when subjected to the MIA. 
To address this vulnerability, we introduce and evaluate the QMU framework, which provides a systematic workflow for effectively revoking the influence of specified data from a trained model. 
Under the same gray-box inference-API setting used in Section \ref{SEC_MIA}, we apply QMU to the deployed target QNN and compare different unlearning mechanisms across noiseless simulation and cloud quantum device, as well as under varying measurement-shot budgets.
This section, therefore, addresses the second core research question: Can machine unlearning enable QML models to efficiently mitigate this leakage?

\begin{figure*}
    \centering
    \includegraphics[width=1\textwidth]{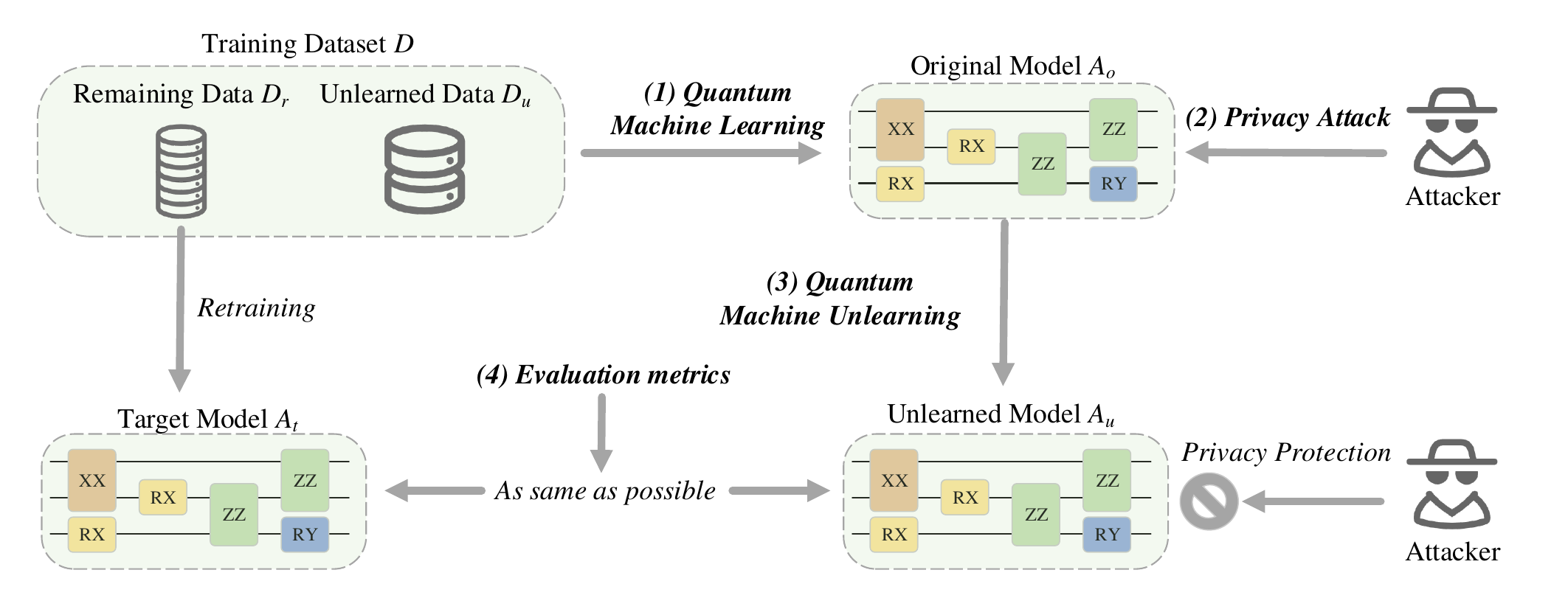}
    \caption{Attack and Unlearning Workflow for QML.
(1) Training: the original QML model $A_o$ is trained on the full dataset $D$; the subset to be revoked is $D_u \subset D$, and the retained data are $D_r = D \setminus D_u$; the target QML model $A_t$ is trained on the dataset $D_r$;
(2) Privacy attack: adversaries launch MIA on $A_o$ to test whether traces of $D_u$ are exposed.  
(3) Unlearning: an algorithm $U$ acts on $A_o$ to produce the unlearned model $A_u$ that should discard information about $D_u$ while preserving performance on $D_r$.  
(4) Evaluating: comparing $A_u$ with the ideal baseline $A_t$ on multiple evaluation metrics, which include accuracy, the success rate of MIA, and computational cost.
}
\label{fig:unlearning}
\end{figure*}

We formalize the QMU workflow as shown in FIG.\ref{fig:unlearning}. 
We first define the full training dataset as $D = {(x, y)}$ where $x$ represents the input data and $y$ the corresponding label. 
This set is partitioned into two disjoint subsets: the unlearned data $D_u \subset D$ whose influence is to be removed, and the retained data $D_r = D \setminus D_u$.
In general,  the size of the unlearned set is much smaller than the retained set, i.e., $|D_u| \ll |D_r|$.
In this work, we focus on class-level unlearning, where all data belonging to a particular class are to be unlearned. 
The QMU workflow is formalized into four distinct processes.
Process (1): An original QML model $A_o$ was trained using the full dataset $D$. To establish a baseline for effective unlearning, we then retrain an initial model solely on the retained subset $D_r$, yielding the target model $A_t$. However, retraining the $A_t$ from scratch is often impractical due to computational cost, data storage limitations, and data accessibility constraints. 
Therefore, $A_t$ is employed as an ideal unlearning baseline model to evaluate alternative unlearning strategies, rather than as a feasible solution in real-world scenarios. 
Process (2): We assess the vulnerability to privacy of the original model $A_o$ by simulating adversarial behavior, such as MIA, to determine whether information about the deleted subset $D_u$ can still be inferred. 
Process (3): Subsequently, an unlearning method $U$ is applied to the original model $A_o$, yielding a modified unlearned model $A_u$=$U(A_o)$, which is designed to eliminate knowledge of $D_u$ and mitigate exposure to privacy attacks. 
Process (4): We propose a set of evaluation metrics to quantify the behavioral similarity between $A_u$ and $A_t$, thereby assessing the effectiveness of the unlearning method.

Overall, the central goal of the QMU workflow is to transform an original model $A_o$, trained on a full dataset, into an unlearned model $A_u$. The unlearned model $A_u$ should effectively forget a specific subset of data $D_u$ while preserving performance on the remaining data $D_r$, thereby emulating a target model $A_t$ that was retrained from scratch, but without the prohibitive computational cost.
To implement the unlearning process, we next introduce and adapt three distinct algorithms designed to effectively erase the influence of specified training samples: (1) Gradient ascent unlearning, which reverses the learning process by maximizing prediction loss; (2) Fisher-based unlearning, which selectively perturbs parameters based on their sensitivity to target samples; and (3) relative gradient ascent, a hybrid approach combining the previous two methods for more controlled unlearning. 
These mechanisms are particularly well-suited for QML due to their direct compatibility with PQCs. The subsequent subsections will introduce each algorithm in detail, alongside the comprehensive metrics used to evaluate their efficacy, performance, privacy robustness, and computational cost.

\subsubsection{Gradient ascent unlearning method}

The traditional learning process operates by minimizing the loss function to improve the model's fit to the training dataset. In the context of machine unlearning, however, the objective is to induce the model to forget specific samples. A natural and direct approach is to reverse the training dynamic: if minimizing the loss corresponds to data memorization, then maximizing the loss represents an active reverse learning process designed to reduce the model's memory of that data. This strategy is known as the gradient ascent (GA) unlearning method. 
Since QNN optimization relies fundamentally on gradient estimation through PQCs, the gradient reversal operation required for GA is directly and efficiently feasible within a QNN model.

The algorithm proceeds as follows. 
(1) Initialization: Set the trained model's parameters as $\theta$ and the unlearned sample as $(x_t, y_t)$; 
(2) Forward prediction: Compute the model output $y'_t = f_{\theta}(x_t)$; 
(3) Loss calculation: Compute the task-related loss function $\mathcal{L}(y'_t, y_t)$, such as cross-entropy for classification tasks; 
(4) Performing gradient ascent to unlearn:
 \begin{equation}
 \theta^{k+1} = \theta^k + \eta \nabla_{\theta{^{k}}} \mathcal{L}(f_{\theta^{k}}(x_t), y_t),
 \label{GA}
 \end{equation}
(5) Repeat steps (2) to (4) until the loss reaches a preset threshold or the prediction confidence falls below a set value, at which point the process stops.
While the GA method effectively reduces the model's fit to the unlearned data, overly aggressive application may degrade performance on the remaining data. The approach's direct applicability to QNN stems from its reliance on efficient PQC gradient computations, although its effectiveness requires carefully balancing the unlearning strength against model preservation.

\subsubsection{Fisher-based unlearning method}

In ML, a model's memory can be quantified through parameter importance analysis, where the Fisher Information Matrix (FIM) serves as a fundamental measure of the output's sensitivity to parameter variations. 
The FIM's core function is equally applicable in QNN, as parameter updates are similarly guided by fitting the training samples. 
This allows the FIM to effectively identify parameters that are particularly sensitive to the unlearned data.
Parameters exhibiting high Fisher information values with respect to specific samples can be interpreted as having explicitly memorized those data points, thereby making them prime targets for selective modification to achieve effective unlearning.
In this work, we implement an efficient Fisher-based unlearning strategy through the application of Selective Synaptic Dampening (SSD) \cite{foster24}. 

The methodology of SSD is structured around two core components: the computation of Fisher information and the subsequent design of the selective modification process.
First, we compute FIM by the second-order gradient of the loss function, and the diagonal elements of FIM represent how parameter variations affect output loss (for a specific sample). While exact computation requires costly second-derivative calculations, we approximate it by the empirical Fisher approximation, which uses squared first gradients. 
For model $A$ and dataset $D_r$ empirical fisher approximation is computed as:
\begin{equation}
F(D) = \mathbb{E}_{(x,y)\in D}\left[\left(\nabla_\theta \log \mathcal{L}(y|x,\theta)\right)^2\right],
\label{fisher}
\end{equation}
where $L$ is the loss function, $\theta$ represents model parameters, and $\nabla_\theta \log L(y|x,\theta)$ denotes the gradient of the log-likelihood, reflecting output sensitivity to parameter changes.
Secondly, SSD selectively perturbs parameters based on their relative importance to $D_u$ versus $D_r$:
\begin{equation}
\label{equ_theta}
\theta_i' =
\begin{cases}
\beta \theta_i, & \text{if } F(D_u)[i] > \alpha F(D_r)[i], \\
\theta_i, & \text{if } F(D_u)[i] \leq \alpha F(D_r)[i],
\end{cases}
\end{equation}
where $i$ indexes parameters and $\alpha$ represent controls selection strictness.
Lastly, SSD applies importance-weighted dampening to targeted parameters:
\begin{equation}
\label{Equ_beta}
\beta = \min\left(\frac{\lambda F(D_u)[i]}{F(D_r)[i]}, 1\right),
\end{equation}
where $\lambda$ represents the protection strength, while $\beta$ adaptively scales the parameter updates. 
The parameter $\alpha$ controls the strictness of parameter selection, defining the proportion of parameters to be disturbed, and is typically set within the range of [0.1,100]. Subsequently, $\lambda$ adaptively scales the parameter updates, where $\lambda$ represents the unlearning or protection strength, generally set between [0.1,5].
This comprehensive mechanism enables the progressive unlearning of $D_u$ while rigorously preserving the model parameters critical for performance on the $D_r$.

\subsubsection{Relative gradient ascent}
To achieve an efficient and highly controlled unlearning method, QMU provides the Relative Gradient Ascent (RGA) method. This approach combines the precision of Fisher information's sensitivity identification with the power of gradient ascent's targeted optimization. 
RGA selectively perturbs only the critical model parameters, allowing for the effective removal of specified sample influence with minimal side effects.
The approach involves three steps: (1) computing FIM $F(D_u)$ and $F(D_r)$ for unlearned data $D_u$ and retained data $D_r$, like Eq.(\ref{fisher}); (2) identifying parameters with relative importance of $D_u$ and $D_r$; and (3) applying selective GA where insignificant parameters are either not updated or updated based on an importance factor, with this work focusing on the masking strategy for computationally efficient precision unlearning.
Here, we introduce the first method, as shown in Eq.(\ref{hy}).
\begin{equation}
\theta_i' =  \theta_i + \eta \nabla_\theta \mathcal{L}(f_\theta(x_t), y_t)  \quad \text{if } F(D_u)[i] > \alpha F(D_r)[i], 
\label{hy}
\end{equation}
where $\theta_i$ is the index $i$ of $\theta$. By performing gradient ascent along the direction of relatively important parameters, this method aims to maximize the unlearning effect at minimal cost while avoiding the degradation of the model's fit to the retained data.

\subsubsection{Evaluation metrics for quantum machine unlearning}
\label{sec3_EM}
To comprehensively evaluate the efficacy and practicality of the proposed QMU framework, we propose a set of four rigorous evaluation metrics that establish a framework for assessing the validity of the QMU method.
The first three metrics are designed to quantify the unlearning objective and effectiveness:
(1) Prediction accuracy of unlearned samples ($Acc_U$): This metric directly measures the intensity of forgetting, where lower values indicate more complete unlearning of the specified data.
(2) Prediction accuracy of remaining samples ($Acc_R$): This evaluates the model's performance on the primary machine learning task, ensuring utility is maintained on the retained dataset.
(3) MIA success rate: This reflects the model's defense capability by quantifying its robustness against privacy attacks.
(4) Computational cost: The measured wall-clock time (in seconds) to assess the computational cost of their unlearning algorithms. 
By focusing on these four metrics, we simultaneously assess the model's success in achieving the core unlearning goals and its practicality in terms of computational resources.

\subsection{Simulation and cloud quantum device demonstrations}

\label{sec4_D}
In this section, we focus on evaluating the effectiveness of QMU mechanisms in both noiseless simulations and on a cloud quantum device.
Specifically, we assess the performance of the three implemented QMU methods, which include the GA, SSD, and RGA unlearning methods, on two representative QNN models (the basic QNN and the HQNN).
The evaluation is rigorously based on the four defined metrics: three metrics quantifying the unlearning objective and computational complexity.
We first conduct an individual performance analysis of each MU method across different scenarios in noiseless simulations. 
Following this comparative investigation, we benchmark the optimal results achieved by each QMU method and subsequently deploy the algorithm onto a cloud quantum device to validate its effectiveness.
Simulation results use analytic evaluations for stable optimization, and QMU demonstrations on a cloud quantum device use $N_{shots}=4096$ per circuit evaluation, consistent with Section \ref{SEC_MIA}.
All results are averaged over 20 random seeds; error bars or confidence intervals are provided where space permits. Computational cost is measured as wall-clock seconds on the corresponding platform; cross-architecture cost is not directly comparable.

\subsubsection{Performance analysis of QMU methods}
\label{QMU_result}
\begin{figure*}
    \centering
    \includegraphics[width=1\textwidth]{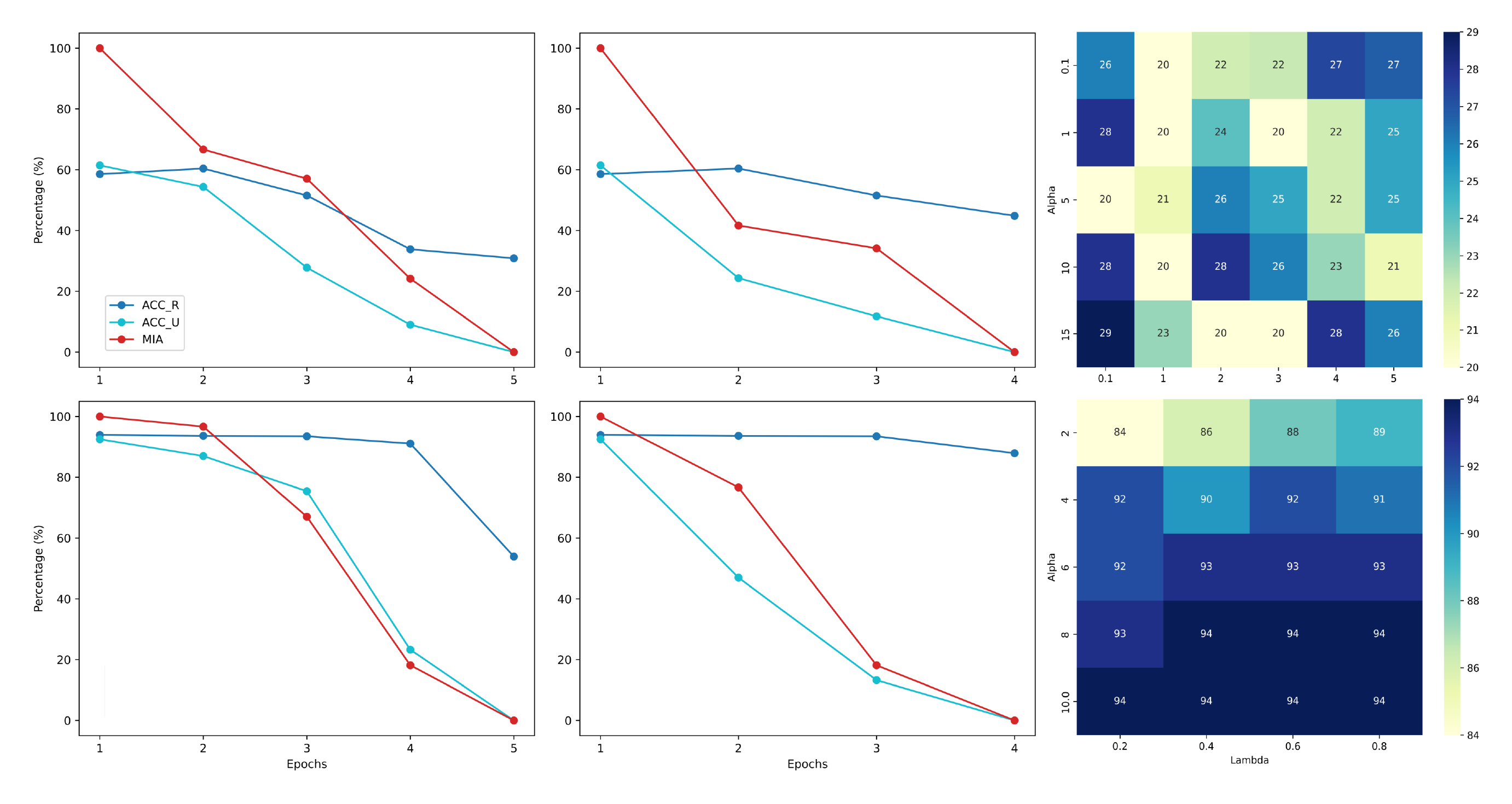}
    \caption{
Performance Comparison of QMU. This figure presents the performance comparison of QMU methods: GA (left), RGA (middle), and SSD (right) on two types of QNN (top) and HQNN (bottom). 
The first two columns display the relationship between the number of epochs and three metrics reported as percentages (\%): classification accuracy on retained data ($ACC_R$) and unlearned data ($ACC_U$), and the MIA success rate (attack accuracy on a member set). 
The heatmaps on the right report the retained-set accuracy $ACC_R$ (\%) achieved by SSD across a grid of $(\alpha,\lambda)$ (defined in Eq.\ref{equ_theta} and \ref{Equ_beta}), where $\alpha$ controls the strictness of parameter selection and $\lambda$ controls the dampening/perturbation strength;
These results demonstrate the trade-offs in effectiveness and efficiency of the different unlearning methods across the two models.
}
\label{fig:QMU_per}
\end{figure*}

In this subsection, we analyze and compare the unlearning effectiveness and utility of the three QMU methods (GA, RGA, and SSD) on both QNN and HQNN. We report the retained-set accuracy $ACC_R$, the unlearned-set accuracy $ACC_U$, and the MIA success rate under the different hyperparameters.
In FIG.\ref{fig:QMU_per}, the y-axis values are reported as percentages (\%). $ACC_R$ and $ACC_U$ denote the classification accuracy on the retained dataset $D_r$ and the unlearned dataset $D_u$, respectively, and the MIA denotes the MIA attack accuracy of member set. 
For SSD, we perform a grid search over $(\alpha,\lambda)$ (see Eq.\ref{equ_theta} and \ref{Equ_beta})) and report the corresponding $ACC_R$ at the setting that minimizes both $ACC_U$ and MIA success rate.

Firstly, we evaluate the performance of the GA method for both QNN architectures (see FIG.\ref{fig:QMU_per}, left).
In the forgetting phase, we implement GA for both QNN architectures with a learning rate of 0.01 and a batch size of 16.
The GA method operates by applying reverse learning specifically to the unlearned dataset $D_u$, making it a highly suitable mechanism for QNN pipelines.
A distinct advantage of GA, compared to the other two methods, is its reliance solely on the $D_u$ dataset, eliminating any dependence on the retained dataset $D_r$. 
This attribute enhances its efficiency in scenarios constrained by data accessibility or limited storage.
In our analysis, GA successfully achieves the unlearning objective on $D_u$. However, it occasionally causes a slight drop in accuracy on $D_r$.
Consequently, model performance can be restored in such scenarios by fine-tuning the model using the $D_r$ dataset following the unlearning procedure. Crucially, GA demonstrates its capability to achieve the primary unlearning objectives across both the basic QNN and HQNN models.

Next, we evaluated the performance of the SSD method (see FIG.\ref{fig:QMU_per}, right). 
SSD operates by selectively perturbing model parameters based on their relative sensitivity to both the unlearned dataset $D_u$ and the retained dataset $D_r$.

Based on our evaluation design and preliminary tests, we set the search ranges for the SSD hyperparameters as follows:
$\alpha \in [0.1, 15]$ and $\lambda \in [0.1, 5]$ for the basic QNN, and
$\alpha \in [2, 10]$ and $\lambda \in [0.2, 0.8]$ for the HQNN.
The results demonstrate that the SSD method performs poorly on the basic QNN model but achieves significantly better outcomes on the HQNN model.
We hypothesize that this discrepancy stems from the intrinsically lower accuracy of the basic QNN, which leads to substantial biases in the parameter importance estimation derived from the FIM.
This bias ultimately compromises the accuracy of the selective noise injection process.
Consequently, these findings suggest that SSD is more effective when applied to models with higher initial accuracy and richer parameter representations, such as the HQNN architecture.

Finally, we evaluate the performance of the RGA method (see FIG.\ref{fig:QMU_per}, middle). 
The RGA method is a hybrid approach, combining the GA optimization strategy with the selective parameter identification provided by FIM.
This combination focuses the gradient updates exclusively on the parameters most sensitive to the unlearned dataset $D_u$.
We set relatively small learning rates to observe the unlearning process of the RGA method.
During the unlearning process of basic QNN, the RGA method was configured with a learning rate of 0.01 and a batch size of 16. 
For HQNN, the RGA baseline was configured with a learning rate of 0.05 and a batch size of 32. 
Unlike GA, RGA requires access to the complete dataset to compute the relative Fisher information, which makes it inherently more computationally demanding.
However, the incorporation of FIM significantly improves the efficiency of the unlearning process by allowing RGA to selectively target the parameters most responsible for memorizing $D_u$.
Our results confirm that RGA successfully unlearns the $D_u$ dataset while consistently maintaining a higher accuracy on the retained dataset $D_r$ compared to other approaches.

\subsubsection{Comparative performance of QMU methods}
\begin{table}[t]
\caption{
Comparison of QMU on noiseless numerical simulation. 
}
    \centering
    \begin{tabular}{l c c c l}
    \toprule
        Method & $Acc_U(\%)$ & $Acc_R(\%)$ & MIA(\%) & Computational cost(s) \\ 
        \toprule
        \multicolumn{5}{c}{Basic QNN} \\
        \toprule
        $A_o$ & 61.2 & 57.1 & 95.1 & 1785.4 \\ 
        $A_t$ & 0 & 64.2 & 0 & 1605.2 \\ 
        \toprule
        GD & 0 & 63.7 & 22.6 & 160.5 \\ 
        \toprule
        GA & 0 & 54.4 & 6.8 & 10.7 \\ 
        GA-R & 0 & 63.5 & 0.0 & 107.1 \\ 
        \toprule
        SSD & 0 & 14.8 & 5.8 & 40.6 \\ 
        SSD-R & 0 & 47.1 & 0 & 92.6 \\ 
        \toprule
        RGA & 1.9 & 59.8 & 2.7 & 48.9 \\
        RGA-R & 0 & 61.4 & 0.0 & 82.3 \\
        \toprule
        \multicolumn{5}{c}{HQNN} \\
        \toprule
        $A_o$ & 96.6 & 94.0 & 100 & 357.5 \\ 
        $A_t$ & 0 & 95.8 & 0.0 & 321.75 \\ 
        \toprule
        GD & 92.0 & 95.1 & 100 & 321.7 \\ 
        \toprule
        GA & 0 & 89.5 & 0.4 & 40.0 \\ 
        GA-R & 0 & 90.7 & 0.0 & 74.2 \\ 
        \toprule
        SSD & 0 & 94.2 & 0.8 & 13.6 \\ 
        \toprule
        RGA & 0 & 93.2 & 0.4 & 42.1 \\ 
        RGA-R & 0 & 96.5 & 0.1 & 76.0 \\ 
    \bottomrule
    \end{tabular}
    \label{table_sim}
\end{table}

In this section, we compare the optimal results achieved by the three proposed MU methods under consistent evaluation settings, using both noiseless numerical simulations and cloud quantum device demonstrations.
We also present the original model $A_o$ and the target model $A_t$ as benchmarks for comparison.
In addition to the QMU, we also deploy the GD method as a comparative baseline. This approach continues training the model solely on the retained dataset $D_r$. 
The intent is that this prolonged training causes the model to overfit to the retained data, thereby causing it to effectively forget the influence of the unlearned dataset $D_u$.
We additionally report optional ”-R” variants, e.g., GA-R, SSD-R, which apply a short fine-tuning on $D_r$ only after the QMU step. This calibration is not required to satisfy the unlearning objective; it is included to examine whether retained-set accuracy can be further improved without degrading forgetting or MIA resistance.
This yields a two-stage procedure in which, after the initial unlearning algorithm successfully reduces the model’s accuracy on the unlearned data to near zero, a brief fine-tuning phase is applied exclusively on $D_r$ to recover model performance. 
All MU methods are rigorously evaluated based on the four key metrics. It is essential to note that because the basic QNN and HQNN models are executed on different platforms, the computational complexity figures for these methods cannot be directly compared across architectures.

The results of the noiseless numerical simulations are shown in Table \ref{table_sim}.
For the GD baseline, the method failed to meet the unlearning objective in the basic QNN due to its high MIA success rate 22.6\%, despite reducing the model's ability to fit the unlearned data.
Furthermore, the GD method proved ineffective for the HQNN.
This suggests that overfitting-based approaches are unsuitable for erasing information associated with $D_u$, primarily because the model's high expressive capacity allows it to retain memorized information even after attempts to unlearn.
During the unlearning process of basic QNN, the GD method was configured with a learning rate of 0.01 and a batch size of 8. For HQNN, the GD method was configured with a learning rate of 0.01 and a batch size of 16. 
In comparison to GD, the proposed QMU methods yielded superior results.
First, the GA method generally achieved the unlearning objective, though it resulted in a slight reduction in accuracy on the retained dataset $D_r$. 
Its primary advantages are the requirement of only the unlearned dataset $D_u$, making it a more feasible option in resource-constrained scenarios.
As observed in Section \ref{QMU_result}, the GA method strongly forgets the target data. Therefore, we set relatively small learning rates of 0.003 for the basic QNN and 0.001 for the HQNN.
Second, the SSD method successfully achieved the unlearning objective in the HQNN, while exhibiting minimal computational complexity, with the hyperparameters set to $(\alpha,\lambda)=(10,0.2)$. However, its major drawback is poor robustness, as it is particularly less effective when applied to models with low accuracy.
Third, the RGA method effectively combines the strengths of GA and SSD. By utilizing the FIM to identify parameters sensitive to $D_u$ and applying GA selectively to modify only those parameters, RGA allows for a larger learning rate and consequently speeds up the unlearning process. 
This approach makes RGA more efficient in achieving complete unlearning without significantly compromising performance on $D_r$.
During the unlearning process of basic QNN, the RGA method was configured with a learning rate of 0.05 and a batch size of 16. For HQNN, the RGA method was configured with a learning rate of 0.1 and a batch size of 16. 
Finally, we also deployed the three MU algorithms on a cloud quantum device and compared their performance. Our results, shown in Table \ref{table_h}, demonstrated performance similar to the trends observed in the noiseless numerical simulations.
Summarizing across the two architectures, GA exhibits the lowest data dependence (uses only 
$D_u$), SSD achieves the lowest cost on the HQNN, and RGA provides the strongest robustness while maintaining $Acc_R$.

\begin{table}[!ht]
\caption{
Comparison of QMU on a cloud quantum device. 
}
\centering
    \begin{tabular}{l l c c c l}
    \toprule
        Method & $Acc_U(\%)$ & $Acc_R(\%)$ & MIA(\%) & Computational cost(s) \\ 
        \toprule
        \multicolumn{5}{c}{Basic QNN} \\
        \toprule
        $A_o$ & 44.9 & 42.2 & 76.7 & 3947.3 \\ 
        $A_t$ & 1.8 & 49.6 & 4.5 & 3566.0 \\ 
        \toprule
        GA-R & 0.1 & 45.2 & 7.7 & 1214.5 \\ 
        RGA-R & 0 & 45.5 & 7.1 & 1082.3 \\
        \toprule
        \multicolumn{5}{c}{HQNN} \\
        \toprule
        $A_o$ & 92.3 & 89.8 & 100 & 2512.3 \\ 
        $A_t$ & 0 & 90.7 & 0.0 & 2486.5 \\ 
        \toprule
        GA & 0 & 86.6 & 1.6 & 812.4 \\ 
        SSD & 0.4 & 88.7 & 1.1 & 413.6 \\ 
        RGA & 0 & 88.3 & 0.9 & 434.6 \\
    \bottomrule
    \end{tabular}
    \label{table_h}
\end{table}

\subsubsection{Analysis of the Impact of shot noise on QMU}

In Section \ref{sec:shot_noise_mia}, we established that shot noise can serve as a passive privacy defense during the inference phase. 
However, for the active process of QMU, this intrinsic physical noise poses a significant challenge. 
QMU algorithms typically rely on precise estimations of model gradients (e.g., GA) or fine-grained quantification of parameter importance (e.g., SSD) to execute targeted data erasure. 
Excessive shot noise can introduce statistical errors that disrupt these precise estimations, potentially causing the optimization process to deviate from its intended trajectory and leading to model utility collapse. Consequently, in this section, we evaluate the robustness of different QMU mechanisms under varying measurement shot counts to quantify the impact of physical noise on unlearning performance.

\begin{figure}[t]
    \centering
    \includegraphics[width=\linewidth]{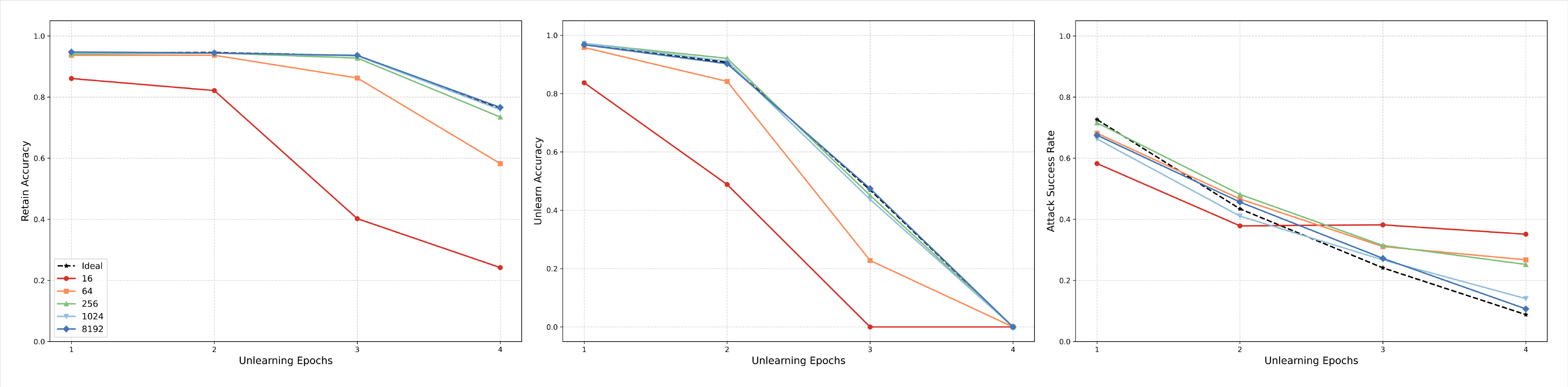} 
    \caption{
    Impact of shot noise on the stability of the GA unlearning process.
    The evolution of performance metrics is tracked over 4 unlearning epochs under varying measurement shot counts. The subplots display: (a) Left: Retain Accuracy ($D_r$), (b) Center: Unlearn Accuracy ($D_u$), and (c) Right: MIA Success Rate. 
    The optimization process exhibits high sensitivity to shot noise. In the low-shot regime (red line, $N_{shots}=16$), the gradient estimation errors lead to a catastrophic collapse in Retain Accuracy, causing the algorithm to degenerate into a stochastic walk away from the optimal parameters. Conversely, high-shot settings ($N_{shots} \ge 1024$, blue/green lines) are required to maintain a stable trajectory converging to the ideal baseline (dashed lines).
    }
    \label{fig:ga_robustness}
\end{figure}

\begin{figure}[t]
    \centering
    \includegraphics[width=\linewidth]{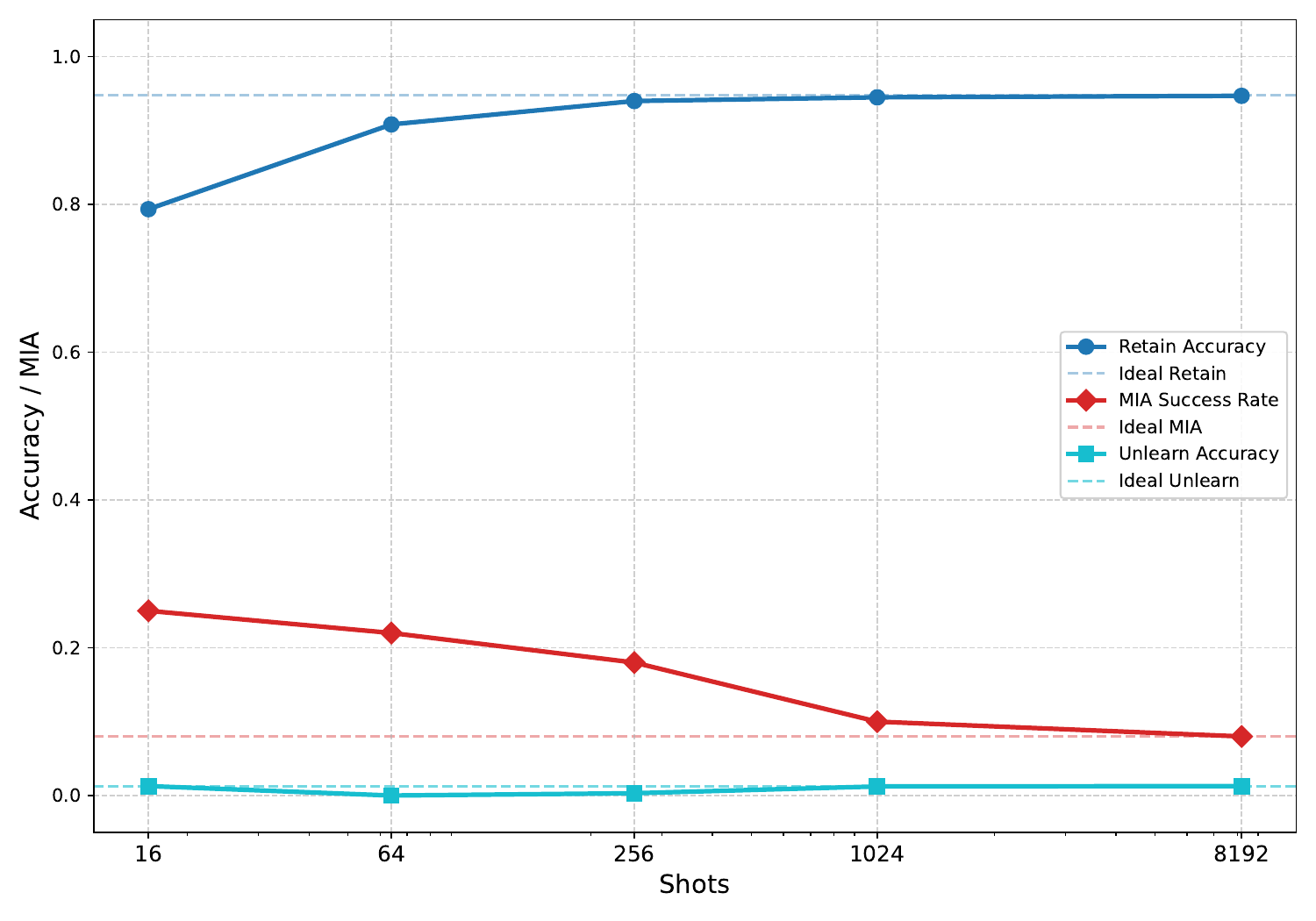} 
    \caption{
    Robustness of the SSD mechanism against shot noise. The plot illustrates post-unlearning performance metrics as a function of measurement shot counts ($N_{shots}$, logarithmic scale). The blue, cyan, and red solid lines represent Retain Accuracy, Unlearn Accuracy, and MIA Success Rate, respectively, while dashed lines indicate their ideal noiseless baselines. 
In stark contrast to GA, SSD demonstrates superior resilience to physical noise. The Retain Accuracy (blue) remains stable ($\sim 0.80$) even in the extreme low-shot regime ($N_{shots}=16$), validating that the relative parameter importance relied upon by SSD is preserved under noise. Furthermore, reducing shots significantly lowers the MIA Success Rate (red), confirming that low-shot inference can serve as a complementary privacy defense without destroying the utility of the SSD-unlearned model.}

    \label{fig:ssd_robustness}
\end{figure}

The simulation results, presented in FIG.\ref{fig:ga_robustness} and FIG.\ref{fig:ssd_robustness}, reveal a fundamental divergence in noise sensitivity between the two algorithms.
The GA algorithm exhibits significant performance degradation as shot count decreases; in the low-shot regime ($N_{\text{shots}}=16$), the retain accuracy suffers a catastrophic decline.
This occurs because GA relies directly on the direction of first-order gradients for its updates, while in PQCs, these gradients are estimated from finite-shot measurements.
Specifically, under the parameter-shift rule, each partial derivative is computed from the difference of two noisy loss estimates,
$\widehat{\mathcal{L}}(\theta_i\pm \pi/2)$, each obtained with $N_{\text{shots}}$ samples, making the gradient estimator unbiased but with variance scaling as $\mathcal{O}(1/N_{\text{shots}})$.
When $N_{\text{shots}}$ is small, sampling noise can dominate the estimated update direction, causing the unlearning trajectory to degenerate into a stochastic walk in parameter space and severely disrupting the previously learned solution, which explains the collapse of $Acc_R$ in FIG.\ref{fig:ga_robustness}.
In contrast, the SSD algorithm demonstrates remarkable robustness in low-shot environments, maintaining stable retain accuracy even at $N_{\text{shots}}=16$.
This stability stems from the second-order nature of SSD: although the Fisher information matrix (FIM) also depends on gradients, SSD primarily leverages the relative sensitivity (ranking) of parameters rather than their exact gradient values.
While shot noise perturbs the FIM estimates, it is less likely to overturn the importance ordering of parameters (i.e., which parameters are most critical).
In short, GA is direction-sensitive (additive shifting), whereas SSD is rank/importance-sensitive (multiplicative dampening), making SSD significantly more tolerant to shot-noise-induced estimation errors.

\subsection{Summary}
This section introduced the QMU framework and established the corresponding evaluation metrics. 
Our evaluation results were systematically conducted to evaluate the applicability of various MU mechanisms across two representative QNN models. The focus of this evaluation was the combined performance in achieving the core unlearning objective and minimizing computational complexity. In conclusion, our findings provide a positive answer to the second research question: with the integration of suitable QMU algorithms, two types of QNN models demonstrably possess the capacity to unlearn training data effectively and with a high degree of control.
Moreover, we analyze the impact of finite-shot measurements on unlearning stability and observe a clear robustness gap across mechanisms, motivating a phase-dependent shot configuration.

\section{Conclusion}\label{sec-con}
This work exposes a concrete membership privacy risk in QNNs and introduces QMU as a practical mitigation strategy.
We first analyze membership privacy leakage in QML under a realistic gray-box inference-API threat model. Under this model, we quantified this privacy leakage using an MIA, demonstrating that trained QNN models leak non-trivial membership information in both noiseless quantum simulators and on a cloud quantum device.
Building upon this finding, we introduced the QMU framework and evaluated its performance using a unified protocol that reports: forgetting strength, retained-set accuracy, MIA success rate, and computational cost.
Our core results indicate that QMU markedly reduces the MIA success rate while successfully preserving high utility on the retained data. 
A comparative analysis further reveals that the three MU mechanisms exhibit distinct trade-offs in data dependence, computational cost, and robustness.
Moreover, our numerical results reveal a two-sided role of measurement shots: a low shot count at inference can mask membership signatures with only a minor accuracy loss, whereas a high shot count improves both attack reliability and unlearning efficacy. Accordingly, we advocate a phase-dependent shot configuration: high shots for maintenance (training/unlearning) and low shots for deployment.

Our study adds a membership privacy and unlearning perspective to the growing QML security literature.
Beyond the present study, an important direction is to deepen the quantum-specific understanding of privacy leakage by characterizing how platform interfaces reshape attacker capabilities and leakage mechanisms.
On the attack side, future work should account for more realistic adversaries under practical constraints.
Future research will prioritize extending the applicability and generalization of existing QMU methods to encompass a broader range of QML models and more diverse datasets.
Currently, unlearning strategies are predominantly concentrated within supervised learning settings, showing limited support for more complex tasks such as unsupervised learning, reinforcement learning, and generative modeling, which warrant thorough exploration. 
Key future directions include the integration of QMU into multi-task learning, secure training workflows, and broader quantum privacy computing frameworks. These methodological extensions are expected to drive quantum learning models towards greater efficiency, controllability, and verifiability, thereby enhancing their practicality and interpretability across a wider range of applications.

\section*{Data Availability}
The code and data supporting the findings of this study are publicly available in Ref.~\cite{qmu_code}.

\section*{Code Availability}
The official implementation is publicly available in Ref.~\cite{qmu_code}.

\section*{Acknowledgments}
This work is supported by the National Natural Science Foundation of China (Grant Nos. 62372048, 62272056, 62371069, and U25B2014).

%

\end{document}